\documentstyle[prc,graphicx,floats,aps]{revtex}
\begin{document}
%\draft
\title{On the temperature dependence of the symmetry energy}

\author{D. J. Dean$^1$, K. Langanke$^2$ and J.M. Sampaio$^2$} 
\address{
$^1$ Physics Division, Oak Ridge National Laboratory, Oak Ridge, TN 37831 USA
\\
$^2$Institut for Fysik og Astronomi, {\AA}rhus Universitet, 
DK-8000 {\AA}rhus C, Denmark
}
\date{\today}
\maketitle

\begin{abstract}
We perform large-scale shell model Monte Carlo (SMMC) calculations 
for many nuclei in the
mass range $A=56-65$ in the complete
$pfg_{9/2}d_{5/2}$ model space using an effective
quadrupole-quadrupole+pairing residual interaction. Our calculations are
performed at finite temperatures between $T=0.33-2$ MeV. Our main focus
is the temperature dependence of the symmetry energy which we 
determine from the energy differences between various isobaric pairs with the
same pairing structure and at different temperatures. 
Our SMMC studies are consistent with an increase of 
the symmetry energy with temperature. We also investigate 
possible consequences for core-collapse supernovae events.
\end{abstract}

\section{Introduction}

Computer simulations of core collapse supernovae are currently at the 
forefront of research in astro-, nuclear and computational physics.
As one-dimensional simulations usually fail to explode 
\cite{Janka01,Mezzacappa01}, the challenging question 
is whether core-collapse supernovae require 
multidimensional effects like convection, 
rotation or magnetic fields, 
or whether the microphysics input into the simulations 
is still insufficient or incorrect. The latter possibility involves 
nuclear physics ingredients like stellar electron 
and $\beta$-decay rates 
\cite{Langanke00,Langanke01,Heger01,Heger01a,Langanke01a}, 
neutrino-nucleus reactions
\cite{Bruenn91,Fuller91,Sampaio01,Sampaio02}, 
neutrino opacities in nuclear matter \cite{Sawyer98,Reddy99} 
and the nuclear equation of state 
(EOS) \cite{Swesty,Sumiyoshi98}. An essential part of the EOS is the symmetry 
energy which describes the energy needed to separate protons and neutrons. 
The special importance of the symmetry energy for the core collapse of a 
massive star stems from the fact that the dynamical evolution of the 
collapse is strongly 
influenced by electron captures on nuclei and free protons. These captures
drive the
matter in the core towards successively more 
neutron-rich nuclei\cite{BBAL,Heger01,Heger01a}.
The symmetry energy impacts these electron
captures in two essential ways. For
finite nuclei, the effectiveness of the 
capture depends on the magnitude
of the symmetry energy $E_s$; i.e. the larger the value of $E_s$ the more 
difficult it is to change protons into neutrons. Furthermore,
via the EOS the symmetry energy influences the amount of free protons
in the core composition   
which in the late stage of the collapse are believed to be the dominant source 
for electron captures \cite{Bruenn85}.

As part of global nuclear mass formulae, the symmetry energy is conventionally
parametrized as $E_s=a_{\rm symm} (N-Z)^2/A$, where $N,Z$ and $A$ are the neutron, charge and mass number of a nucleus. The value of the coefficient 
$a_{\rm symm}$ can be determined by fit to the observed nuclear masses.
For the application of the symmetry energy in core-collapse supernovae
simulations, Donati {\it et al.} pointed to a possible temperature 
dependence of the $a_{\rm symm}$ coefficient \cite{Donati} which at the 
finite temperatures of the stellar collapse has been estimated to be
somewhat larger than for stable nuclei ($T=0$) hindering electron captures.
The hypothesis is based on the following chain of arguments \cite{Donati}.
On the mean-field level correlations are successfully accounted for by 
using an effective mass parameter $m^\star = m_{\rm \omega} m_k / m$ rather than
the bare nucleon mass $m$. Here, the k-mass $m_k$ simulates spatial 
non-localities in the Hartree-Fock potential introduced by the 
Pauli principle. Furthermore, the ground state mean field is found to be
highly dynamical with strong coupling to surface vibrations. The non-localities
associated with these surface couplings can be absorbed into the so-called
$\omega$-mass $m_\omega$. This mass is found to be $m_\omega \sim 1.5$ m.
However, as it is related to vibrations at low excitation energies of a few
MeV, Donati {\it et al.} pointed out that $m_\omega$ should vary over
the temperature range involved in the collapse phase of a supernova
$(T \le 2$ MeV). On the other hand, the typical energy scale of $m_k$ is 
$\approx 8$ MeV so that variations of $m_k$ with temperature 
are assumed to be small
and unimportant during the collapse \cite{Donati}.
Within the Fermi gas model, 
the symmetry energy $E_s$ depends, 
via its kinetic part, on the change
of the $\omega$-mass 
which scales like $\sim 1/m^\star$.
Hence, a reduction of $m_\omega$ with temperature increases the symmetry energy.
Donati {\it et al.} supported their arguments
by calculations of selected even-even nuclei at finite temperature using the framework of the quasiparticle 
random phase approximation (QRPA). 

Clearly nuclear excitations at low energies also depend on 
correlations beyond those treated on the QRPA level. 
The model of choice for a realistic
description of these correlations (e.g. pairing, 2-particle-2-hole and 
higher order correlations) is the nuclear shell model.
Until recently, such shell model calculations for the nuclei involved in
core-collapse supernovae were impractical due to the prohibitively
large model spaces. This shortcoming has been overcome by modern shell model
developments \cite{Johnson93,Caurier00,Otsuka}, where the Shell Model Monte 
Carlo (SMMC) technique also allows such large-scale calculations at finite
temperature \cite{Dean94}. A first attempt to study the temperature
dependence of the symmetry energy within the SMMC has been reported in
\cite{Dean95}, finding no evidence for an increase of $E_s$ with temperature.
However, this calculation, performed for nuclei with $A\sim60$ using the
KB3 residual interaction \cite{Poves81}, has been made more difficult by the
`g-extrapolation procedure' \cite{Alhassid94} required to circumvent the
notorious sign-problem in SMMC studies with realistic interactions. 
Furthermore, the model space considered (full $pf$ shell) was probably
too small
for calculations at temperatures higher than about $T=1$ MeV. 
We were thus motivated to perform improved 
SMMC studies of the temperature dependence of $E_s$. To this end, we
adopted a significantly larger model space, now also including the
$g_{9/2}$ and $d_{5/2}$ orbitals; these orbitals become important especially as $A$ becomes larger than 65. We also adopted a schematical
but still reasonable residual interaction of the pairing+quadrupole
type. Such an interaction does not give rise to the sign problem
\cite{Lang94} and allows for a determination of observables at finite
temperature without employing the g-extrapolation, thus 
reducing the statistical uncertainties of the results.

\section{SMMC calculations of the symmetry energy}

The SMMC method has been presented in great detail in \cite{report}. 
The SMMC technique
describes nuclear observables 
$\langle A \rangle$ at finite temperature as thermal
averages
\begin{equation}
\langle A \rangle = \frac{{\rm Tr}_A (A e^{-\beta H})}{{\rm Tr}_A e^{-\beta H}}
\end{equation}
where $\beta=1/T$, ${\rm Tr}_A$ denotes the many-body 
trace at fixed particle number $A$ (in this case at fixed 
neutron and proton numbers) and $e^{-\beta H}$ the
many-body propagator. Applying the Hubbard-Stratonovich transformation, 
the components of 
$e^{-\beta H}$ stemming from the two-body parts of the Hamiltonian 
(related to the residual interaction) are transformed into integrals
over many one-body propagators involving fluctuating external fields. 
The necessary integrations
are performed by Monte Carlo sampling techniques.  For certain classes of 
residual interactions
like the attractive pairing+quadrupole force employed here, 
the evaluation of $\langle A \rangle$ is exact,
subject only to statistical errors related to the Monte Carlo integrations.

As mentioned above, our SMMC calculations are performed within the complete
$(pfg_{9/2}d_{5/2})$ model space. The single particle energies have been 
determined from a Woods-Saxon potential parametrization of $^{56}$Ni. As we
are concerned here with a description of collective correlations 
at low energies,
we have employed a pairing+quadrupole-quadrupole Hamiltonian
\begin{equation}
H =\sum_{jmt_z}\epsilon(j)a_{jmt_z}^\dagger a_{jmt_z}-
\frac{G}{4} \sum_{\alpha,\alpha',t_z}
P^\dagger_{JT=01,t_z}(\alpha) P_{JT=01,t_z}(\alpha')
-\chi\sum_{\mu} (-1)^{\mu}Q_{2\mu}Q_{2-\mu}
\end{equation}
where $Q_{2\mu}$ is the mass quadrupole moment operator given by
\begin{equation}
Q_{2\mu} = \frac{1}{\sqrt{5}}\sum_{ab} \langle j_a \mid\mid \frac{dV}{dr}Y_2
\mid\mid j_b \rangle \left[a_{j_a}^\dagger \times a_{j_b}^\dagger\right]^{2\mu}
\end{equation}
with projection
$\mu$ and $a_{jmt_z}^\dagger$ ($a_{jmt_z}$) 
creates (destroys) a nucleon of isospin
projection $t_z$ in the orbital $jm$, 
and the pairing operator $P^{\dagger}$
is defined as
\begin{equation}
P_{JT=01,t_z}^\dagger(\alpha)  = 
(-)^{l} \left[a^\dagger_{\alpha,t_z}
\times a^\dagger_{\alpha,t_z}\right]^{JM=00,T=1} \;,
\end{equation}
with $\alpha=\left\{n,l,j\right\}$.

The one-body potential $V$ is of the Woods-Saxon form \cite{alh96}. 
We averaged contributions of the proton and neutron radial integrals
to the quadrupole reduced matrix elements. 
Using
the parameters $G=0.212$~MeV and $\chi=0.0104$~MeV$^{-1}$fm$^2$, we 
reproduce the collective spectrum in $^{64}$Ni and $^{64}$Ge.
We have checked that center-of-mass contaminations are very small and 
do not affect our results. Our SMMC calculations have been performed with
4096 statistical samples, splitting the temperature parameter $\beta$ into
$\beta = N_\beta\Delta\beta$ slices with $\Delta \beta = 1/32$ MeV$^{-1}$
(for details see \cite{report}).

Besides collective excitations, nuclei in the mass range $A\sim 55-65$, as
encountered in the early stage of the core collapse, 
can be strongly influenced
by pairing and shell correlations due to the vicinity of the $N=Z=28$ 
magic number. All of these modes are governed by energy scales of order
$1-2$ MeV, and hence are expected to change in the temperature range of 
interest. To explore the influence of the various correlations on
the nuclear properties and their temperature dependence, we have performed
SMMC calculations for the $A=56$ and $A=59$ isobars in the temperature
range $T=0.33-2$ MeV.

\begin{figure}
\begin{center}
  \includegraphics[width=0.9\columnwidth]{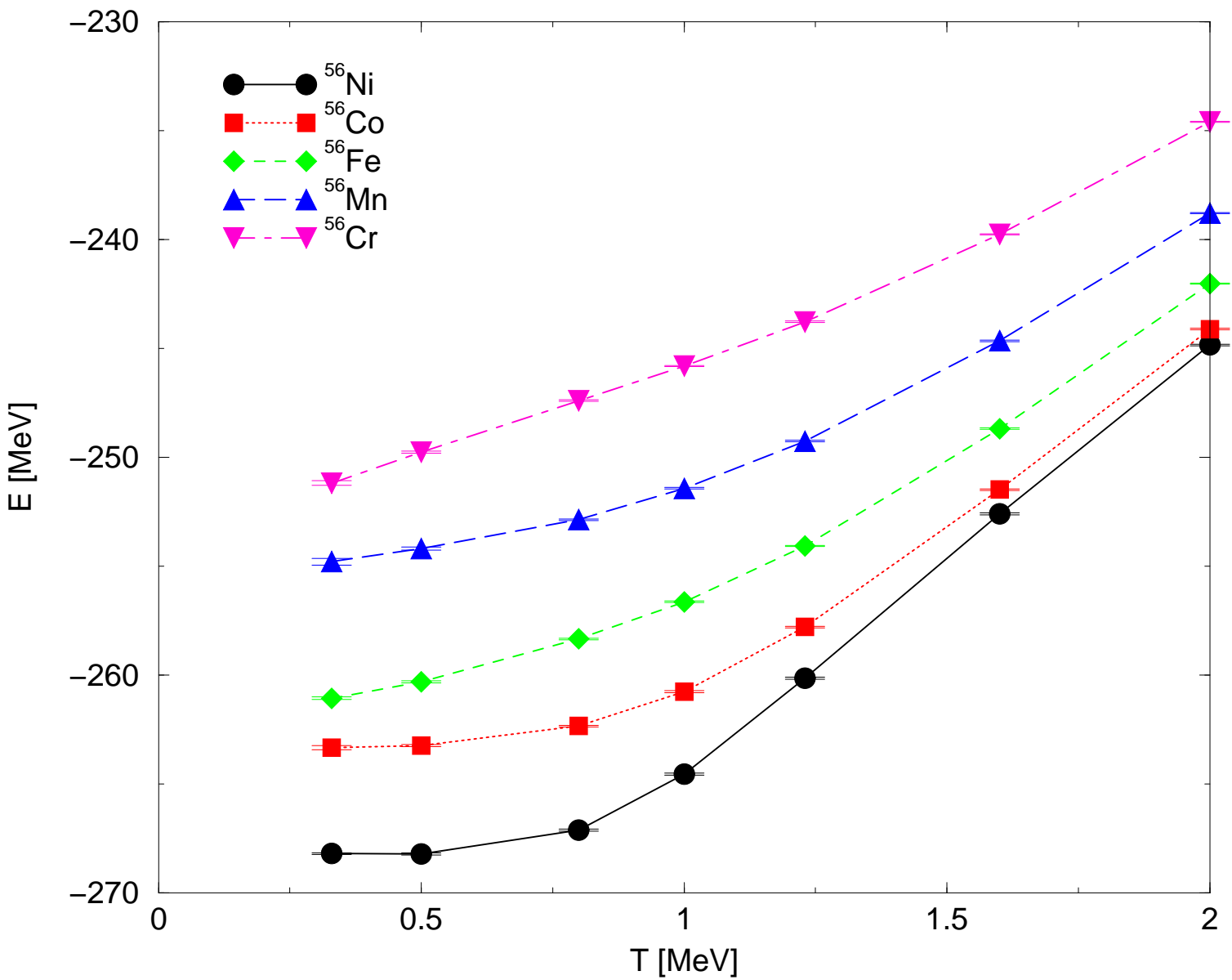}
\end{center}
\caption{
Energy expectation value $E=\langle H \rangle$ as a function of temperature
$T$ for selected $A=56$ isobars.
}
\label{a56_1}
\end{figure}

\begin{figure}
\begin{center}
  \includegraphics[width=0.9\columnwidth]{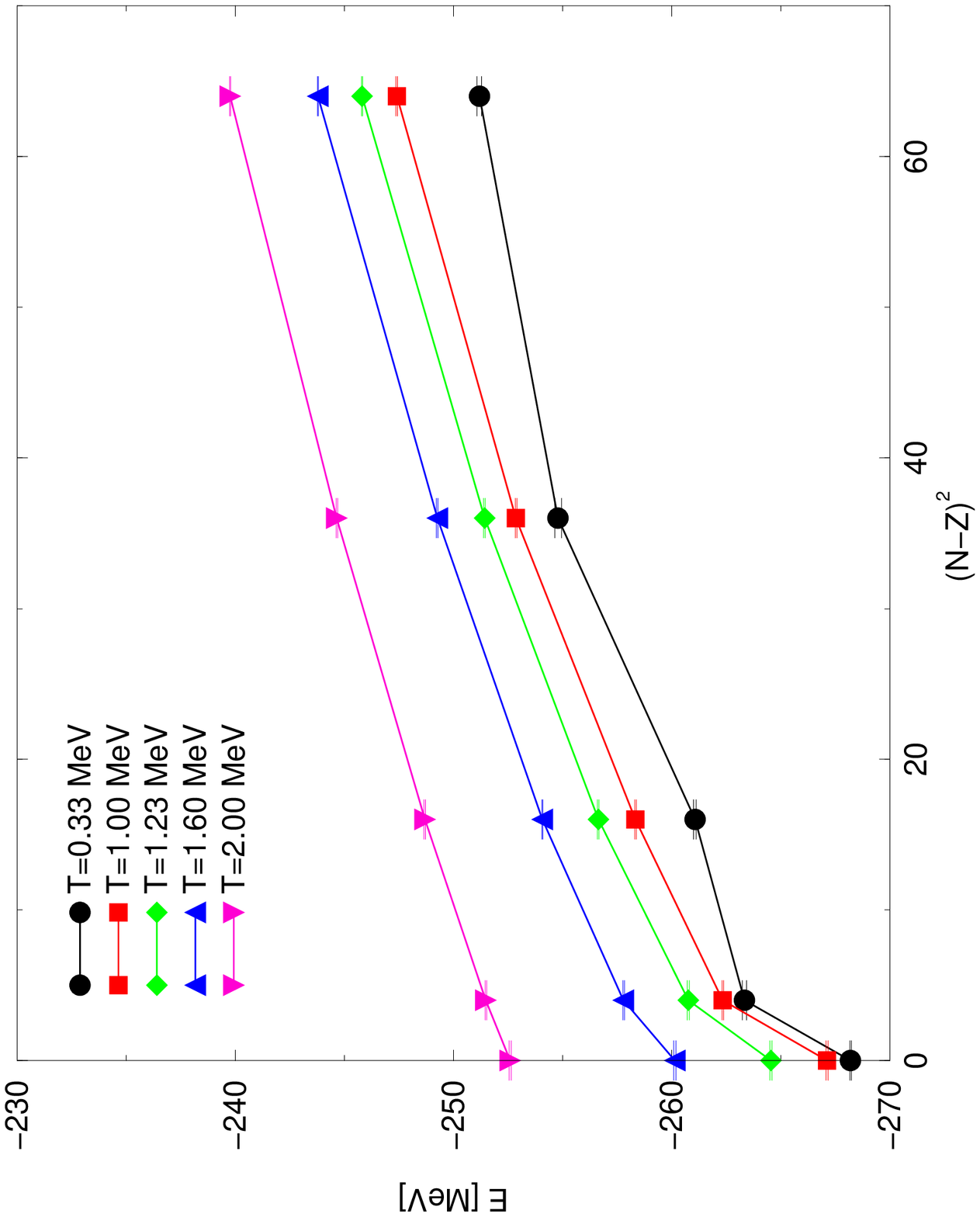}
\end{center}
\caption{
Energy expectation value $E(T)$ as a function of neutron excess
calculated at different temperatures
for selected $A=56$ isobars.
}
\label{a56_2}
\end{figure}

Fig. \ref{a56_1} shows the energy expectation values $E(T)$ as a function of 
temperature, calculated for the even-even ($^{56}$Ni, $^{56}$Fe, $^{56}$Cr)
and odd-odd ($^{56}$Co, $^{56}$Mn) $A=56$ isobars. Due to the presence of
correlations, $E(T)$ exhibits strong deviations at low temperatures from the
$E(T) \sim T^2$ scaling as expected from the simple Fermi gas model.
The large gaps between $^{56}$Ni-$^{56}$Co and $^{56}$Fe-$^{56}$Mn are caused by
the differences in pairing energy between even-even and odd-odd nuclei.
The vicinity of the $N=28$ shell gap dominates the low-temperature
behavior, in particular for $^{56}$Ni and $^{56}$Co. 
This is more clearly
visible in Fig. \ref{a56_2}, where $E(T)$ is plotted against $(N-Z)^2$. In the
conventional Bethe-Weizs\"acker parametrization one has $E \sim (N-Z)^2$
in the absence of pairing and shell correlations (and without consideration
of the Coulomb interaction which we also ignore). Our results suggest
that the pairing correlations are approximately overcome at $T\sim 1.2$ MeV, 
while the shell correlations persist to somewhat higher temperatures.

\begin{figure}
\begin{center}
  \includegraphics[width=0.9\columnwidth]{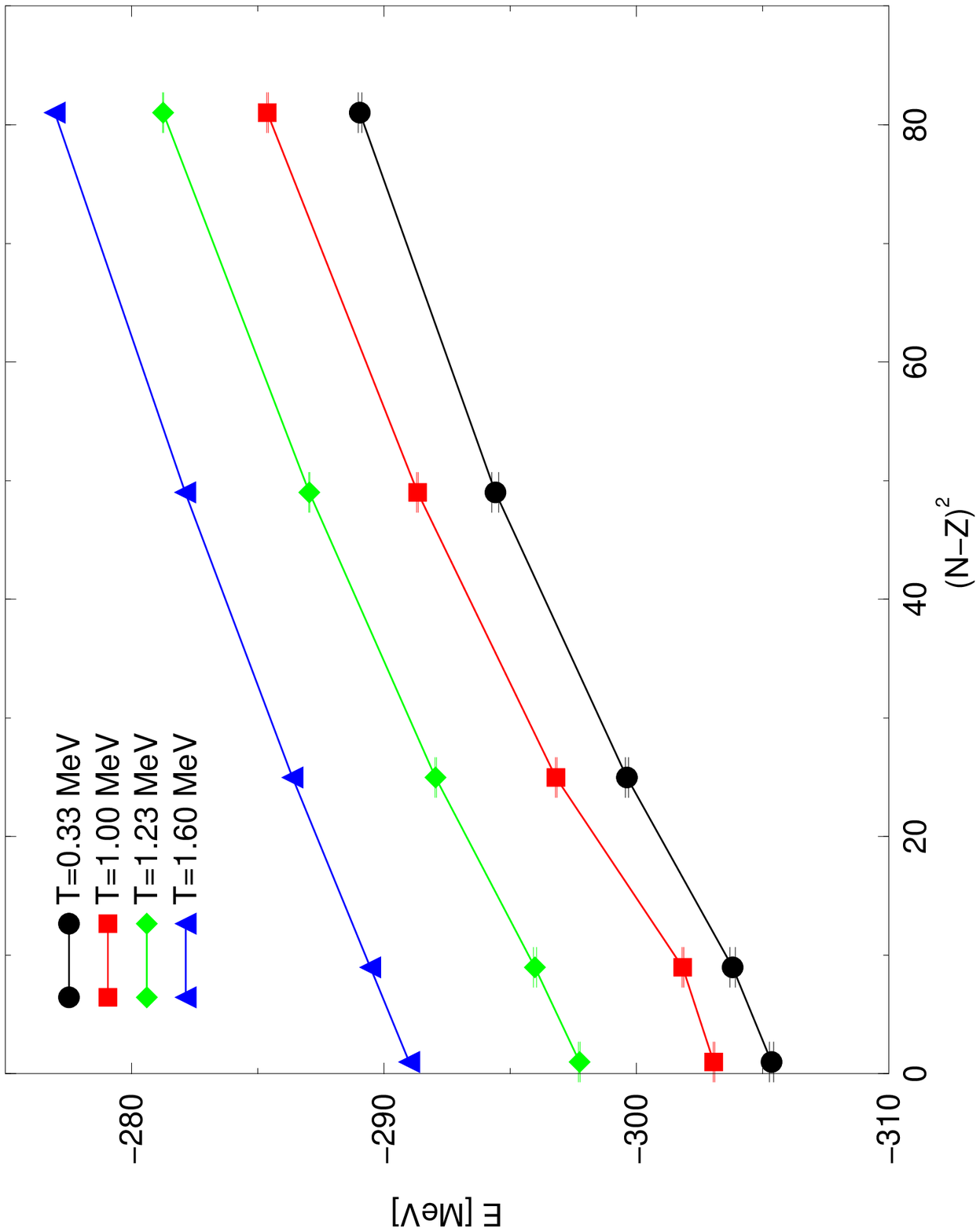}
\end{center}
\caption{
Energy expectation value $E(T)$ as a function of neutron excess
calculated at different temperatures
for selected $A=59$ isobars.
}
\label{a59}
\end{figure}

Fig. \ref{a59} shows the energy $E(T)$ as a function of $(N-Z)^2$ for the 
$A=59$ isobars ($^{59}$Cu, $^{59}$Ni, $^{59}$Co, $^{59}$Fe, $^{59}$Mn).
As expected for odd-$A$ nuclei, the virtual absence of differences
in the pairing energies results in an approximate $(N-Z)^2$ scaling,
already at low temperatures. The presence of the $N=28$ shell gap is responsible
for the deviations from the $(N-Z)^2$ behavior, particularly for $^{59}$Co
and $^{59}$Ni, at low temperatures.

\begin{figure}
\begin{center}
  \includegraphics[width=0.9\columnwidth]{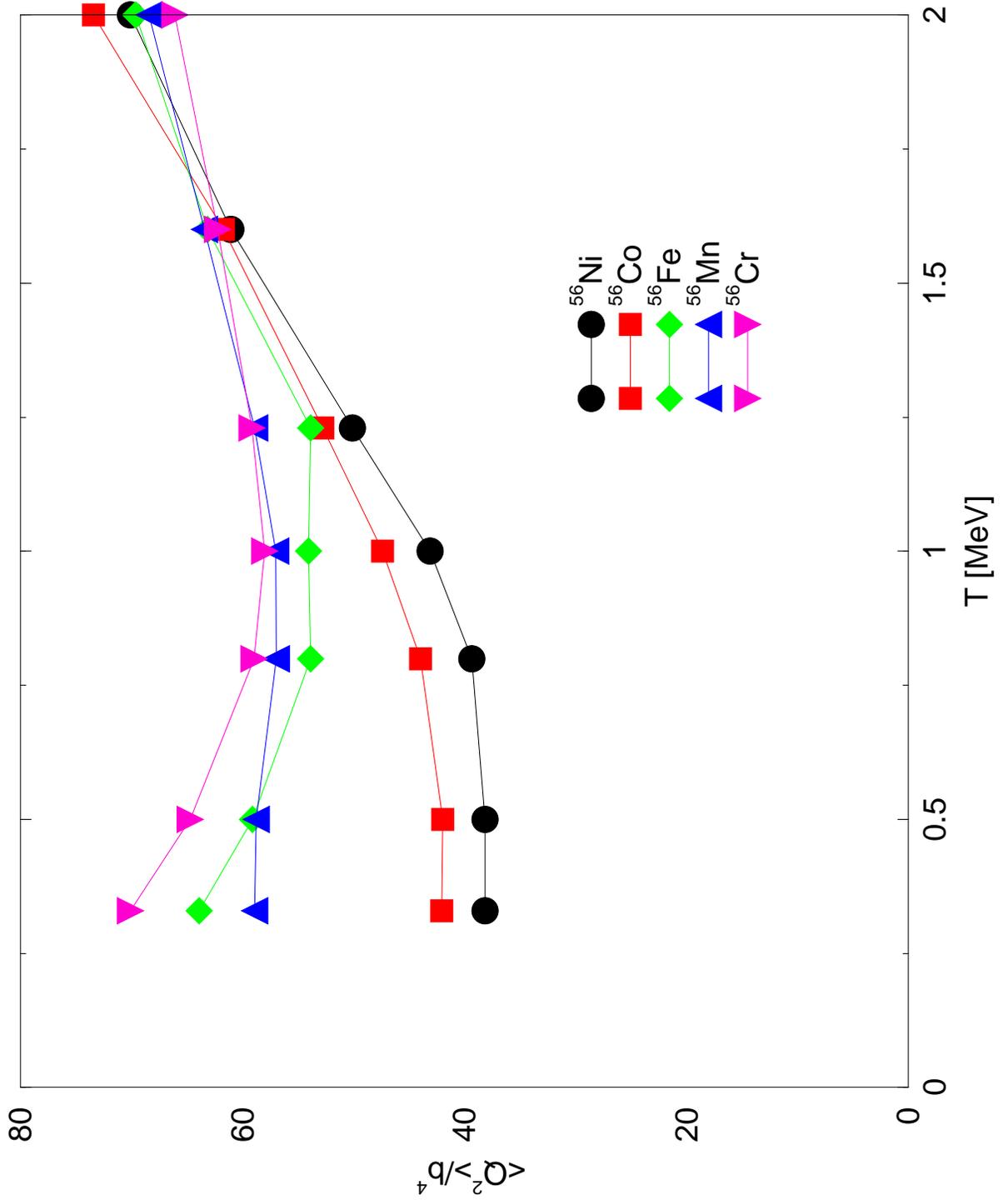}
\end{center}
\caption{
Expectation value of the isovector mass quadrupole 
$\langle Q^2\rangle$ as a function of temperature
for selected $A=56$ isobars.
}
\label{a56_q}
\end{figure}

We show in Fig.~\ref{a56_q} the
expectation value of the isoscalar mass
quadrupole moment $\langle Q^2 \rangle$ as a function of temperature.
The $0f_{7/2}$ shell closure ($N=Z=28$) makes 
the nuclei $^{56}$Ni and $^{56}$Co
spherical and reduces the mass quadrupole at 
low temperatures in this vicinity of
the nuclear chart. For
these two nuclei the shell correlations are overcome with increasing $T$ and
$\langle Q^2 \rangle$ rises. This is different for $^{56}$Fe and $^{56}$Cr,
which are quite collective in the ground state. With the breaking of the pairs
(mainly of identical nucleons), caused by rising temperature, the
mass quadrupole is reduced up to $T \sim 1$ MeV. Due to 
the presence of an unpaired
proton and neutron, the change of the mass quadrupole
is rather moderate in $^{56}$Mn.
At higher temperatures ($T \ge 1.2$ MeV) the nuclei are approximately 
described by uncorrelated nucleons. The mass quadrupole increases as more
nucleons are thermally promoted into the $g_{9/2}$ orbital.

\begin{figure}
\begin{center}
  \includegraphics[width=0.9\columnwidth]{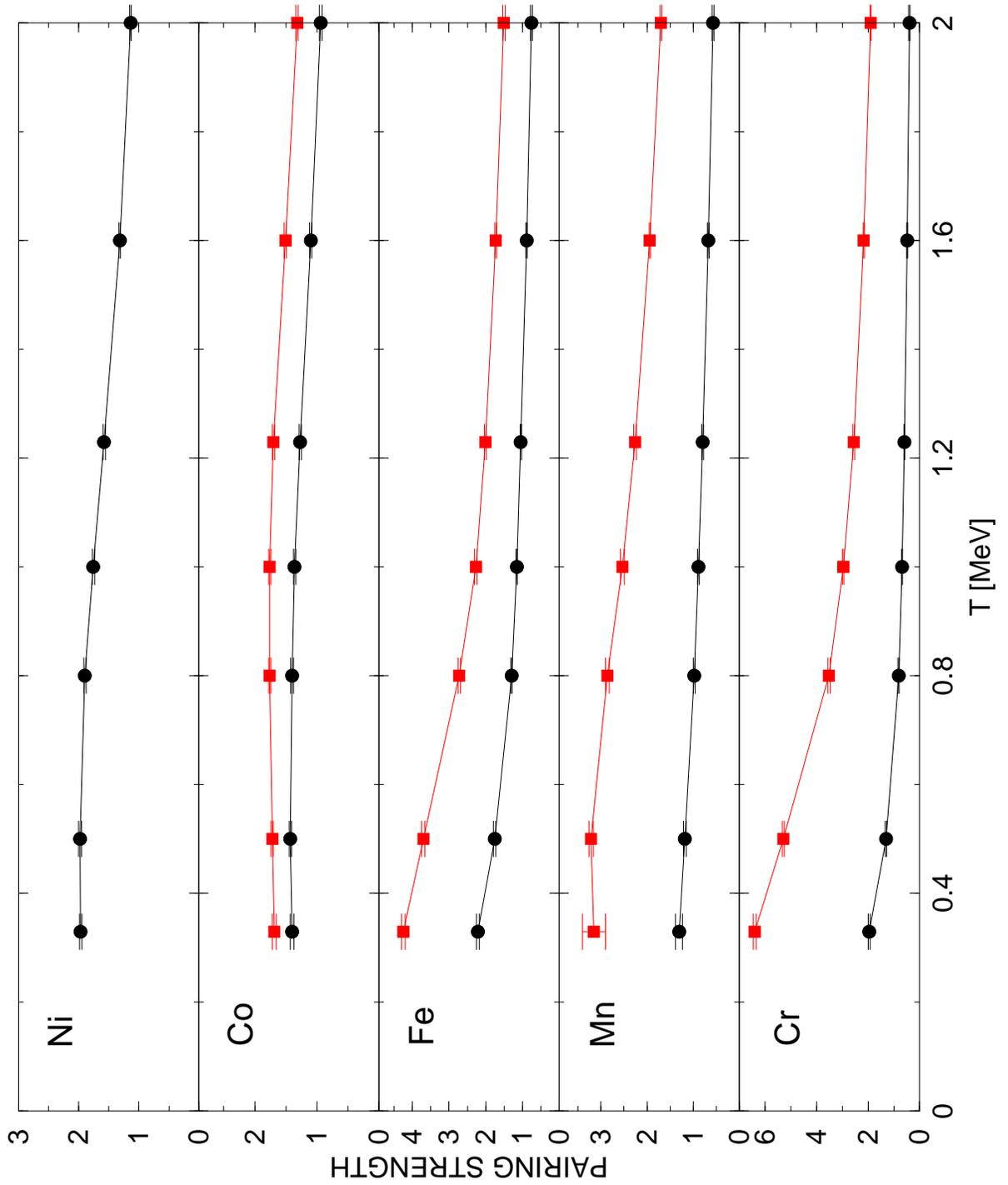}
\end{center}
\caption{
Isovector proton (full circles) and neutron (squares)
pairing strength $P(J=0)$ 
as a function of temperature
for selected $A=56$ isobars.
}
\label{a56_p}
\end{figure}

In Fig. \ref{a56_p} we plot the 
$J=0$ pairing strength between identical
nucleons as a function of temperature for the
set of $A=56$ isobars. As in \cite{Langanke96} we define the 
pairing strength as 
\begin{equation}
P(J=0) = \sum_{\alpha \ge \alpha^\prime} 
M_{\alpha \alpha^\prime}^{J=0}
\end{equation}
with the pair matrix
\begin{equation}
M_{\alpha \alpha^\prime}^{J=0}=
\langle 
A_{J=0}^{\dagger }(j_a,j_b) 
A_{J=0}(j_c,j_d)
\rangle
\end{equation}
where $\alpha=(j_a,j_b)$ and $\alpha^\prime=(j_c,j_d)$.
With increasing temperature the pairing strength
decreases to the mean-field expectation value which, with our definition,
is non-zero \cite{Langanke96}. 

For the $N=Z$ nucleus $^{56}$Ni the proton and neutron pairing strength are
identical. Due to the magicity of this nucleus, pairing is strongly reduced
even in the ground state and is only slightly larger than the mean-field
value at higher temperatures. The even-even nuclei $^{56}$Fe and $^{56}$Cr
exhibit strong pairing among identical nucleons in the ground state. This 
pairing decreases rather strongly with increasing temperature. In the 
odd-odd nuclei ($^{56}$Co, $^{56}$Mn) the unpaired nucleons block pairing
correlations. As a consequence, for example,
the proton pairing strengths in these nuclei
are smaller than in the neighboring nuclei $^{56}$Ni and $^{56}$Fe, 
despite the larger number of valence protons.
In the high-$T$ (mean field) limit the proton and 
neutron pairing strength obviously increase 
with the number of valence protons and neutrons.

We have shown in previous paragraphs a basic description of the structural
properties of our nuclei as a function of temperature. We now turn
to a calculation of the temperature dependence 
of the symmetry energy using our SMMC results. 
In the absence of pairing and shell
correlations, one expects the energy expectation value of isobars,
i.e. for nuclei with constant $N+Z$, to scale like $(N-Z)^2$. To minimize
the differences in pairing and shell correlation energies we choose
several isobaric pairs with the same pairing structure (i.e. even-even, 
odd-odd, odd-$A$) and mainly involving nuclei off the shell closure 
at $N=Z=28$. Our 
pairs are ($^{56}$Fe, $^{56}$Cr), ($^{59}$Fe, $^{59}$Mn),
($^{59}$Cu, $^{59}$Ni), ($^{61}$Cu, $^{61}$Ga), ($^{64}$Ga, $^{64}$Cu),
($^{64}$Zn, $^{64}$Ni), ($^{65}$Cu, $^{65}$Ga), ($^{66}$Zn, $^{66}$Ni),
($^{63}$Ga, $^{63}$Cu). For each nucleus (identified by charge and neutron 
number) we have calculated the energy expectation value $E(T,Z,N)$ at 
temperatures $T=0.33$ MeV and 1.23 MeV. The calculations at the lower
temperature, at least for even-even nuclei, resemble the ground state
properties rather well. The second temperature has been chosen as a
compromise such that $T$ is, on one hand, large enough compared to
the energy scales of the correlations, but on the other hand, small enough
not to be limited by the large, but still restricted, model space.
With the abbreviations $U (T) = E(T,Z_1,N_1) - E(T,Z_2,N_2)$ and
$\Delta [ (N-Z)^2 ] = (N_1-Z_1)^2 - (N_2-Z_2)^2$ we define the
asymmetry parameter
\begin{equation}
a_{\rm symm} (T) = \frac{A}{\Delta [(N-Z)^2]}U(T)\;.
\end{equation}

\begin{figure}
\begin{center}
  \includegraphics[width=0.9\columnwidth]{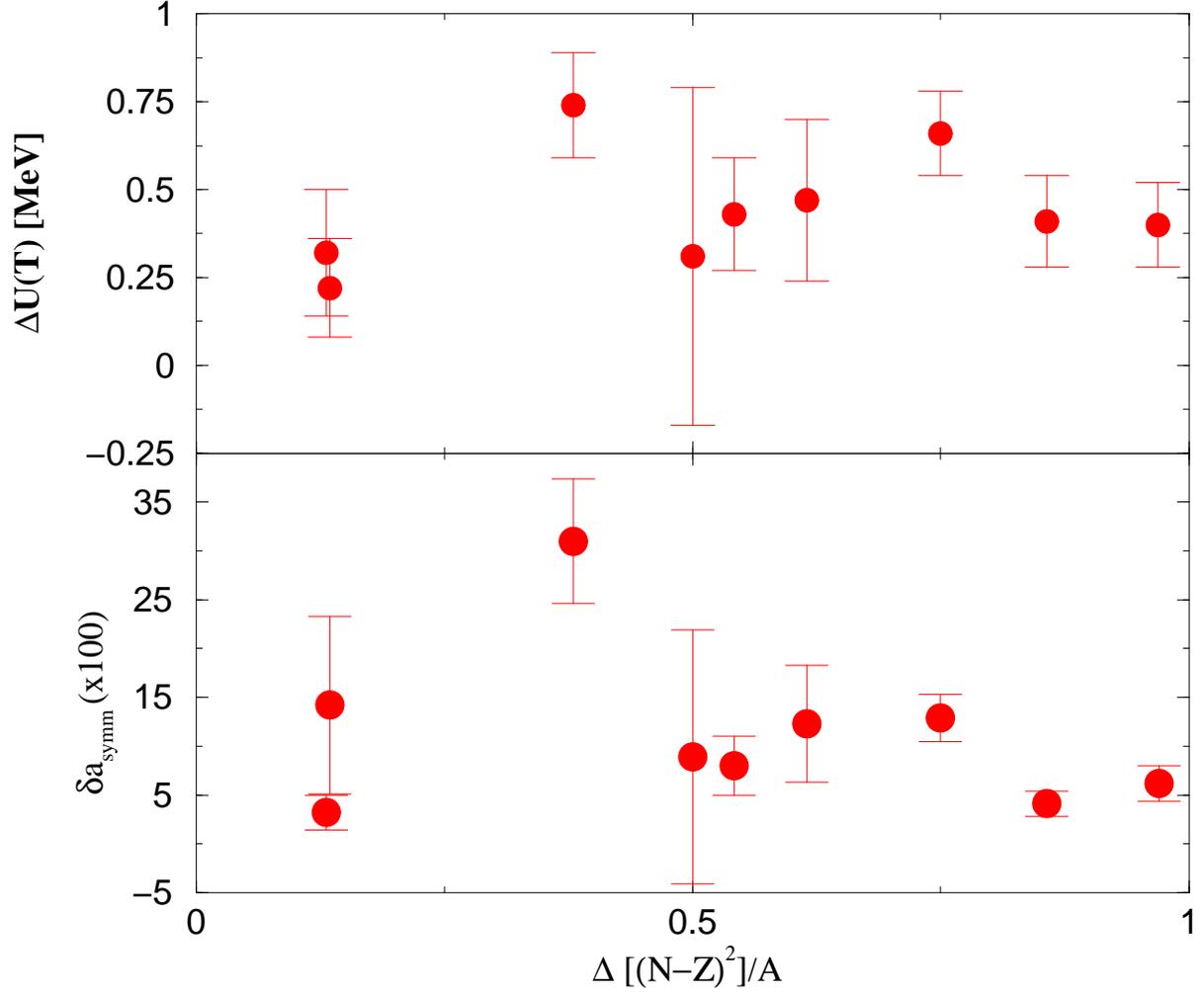}
\end{center}
\caption{
$\Delta U(T)$ (upper panel) and $\delta a_{\rm symm}$ (lower panel)
for the  pairs of isobars specified in the text and identified
by the quantity
$\Delta (N-Z)^2/A$. The single large error bar comes from taking energy
differences between odd-A nuclei.  Generally odd-A systems
possess a larger statistical error for a given number of samples than
their even-even or odd-A counterparts. 
}
\label{symm}
\end{figure}

At $T=0.33$ MeV our calculation yields values for $a_{\rm symm}$ between
7-12 MeV which is smaller than the value determined from experimental
masses, $a_{\rm symm} =17$ MeV, indicating shortcomings in 
the isospin dependence of our schematic residual
interaction. To study the temperature dependence of the symmetry 
energy, we 
are interested in the relative change of $a_{\rm symm}$ with temperature as
\begin{equation}
\delta a_{\rm symm} (T) = \frac{a_{\rm symm} (T) - a_{\rm symm} (T_0)}
{a_{\rm symm} (T_0)}
\end{equation}
where $T_0=0.33$ MeV and $T=1.23$ MeV. (Calculations at even lower temperatures
than 0.33 MeV became numerically unstable in our large model space
due to the relative scale of the single-particle matrix
elements involved in these systems.)
Fig. \ref{symm} summarizes $\Delta U(T)=U(T)-U(T_0)$ and $\delta a_{\rm symm}$ for our 9
isobaric pairs. For all pairs our calculations are compatible with
$\Delta U(T) \ge 0$, i.e. the symmetry energy increases with 
temperature, as predicted by Donati {\it et al.}. Upon averaging over
the $\delta a_{\rm symm}$ for the various pairs, we find 
${\bar \delta a_{\rm symm}}=(6.2 \pm 1.8) \%$ which is approximately in
agreement with the QRPA result of \cite{Donati} which finds an increase
of $\sim 8\%$ of the symmetry energy between $T=0$ and $T=1$ MeV.

\section{Discussion}

Our SMMC calculations are consistent with an increase of the
symmetry energy with temperature, supporting the argumentation of Donati
{\it et al.}. In concluding, we explore the possible consequences 
for the supernova collapse. As pointed out in \cite{Donati} the 
larger symmetry energy at
finite temperature will result in less electron captures and larger
homologous cores, provided all other inputs are kept unchanged. However,
here we like to add a word of caution.
At first, we note that the recent stellar weak
interaction rates \cite{Langanke00,Langanke01} are based on
shell model calculations performed in model spaces large enough to describe
nuclear correlations and their changes with temperature properly for the
temperature regime relevant for presupernova simulations
\cite{Heger01,Heger01a}. These calculations described the stellar
weak-interaction processes for nuclei up to mass $A=65$ as they dominate
the mass composition in the core of a collapsing star up to densities of
order $10^{10}$ g/cm$^3$. As the continuous electron captures drive the
matter more neutron-rich during the collapse, nuclei heavier than $A=65$
become important and even dominate at higher densities. For these nuclei,
diagonalization shell model calculations similar to those performed in
\cite{Langanke00,Langanke01}, are not feasible due to computational
restrictions. To evaluate the relevant electron capture rates for these
nuclei, a hybrid model has recently been suggested \cite{Langanke01a}. In
this model the capture rates are calculated within an RPA approach with
partial occupation formalism, including allowed and forbidden
transitions. The partial occupation numbers represent an `average' state
of the parent nucleus and depend on temperature. They are calculated
within our current SMMC approach at finite temperature. In the hybrid
model the difference of neutron $(\mu_n)$ and 
proton $(\mu_p)$ chemical potentials are fixed
by the $Q$-value of the capture reaction, assuming that there is no
temperature dependence in the symmetry energy. The temperature
dependence of the symmetry energy, however, should slightly increase the
$\mu_n-\mu_p$ difference. To study its relevance for the electron
capture rates during the supernova collapse, we have calculated
rates for three nuclei ($^{72}$Zn, $^{76}$Ga, $^{93}$Kr) which are atypical
representatives of the heavy nuclei at densities $\sim 10^{10}$ g/cm$^3$
and a few $10^{11}$ g/cm$^3$ ($^{93}$Kr) \cite{Liebendoerfer}. The calculations
have been performed for $\mu_n-\mu_p=Q$, i.e. for no temperature dependence
of the symmetry energy, and increasing the $\mu_n-\mu_p$ difference
according to the parametrized increase of the symmetry energy 
given in \cite{Donati}. Possible final-state blocking by neutrinos have
been ignored, although this effect becomes increasingly important in
actual collapse simulations at densities in excess of order $10^{11}$
g/cm$^3$.

\begin{figure}
\begin{center}
  \includegraphics[width=0.9\columnwidth]{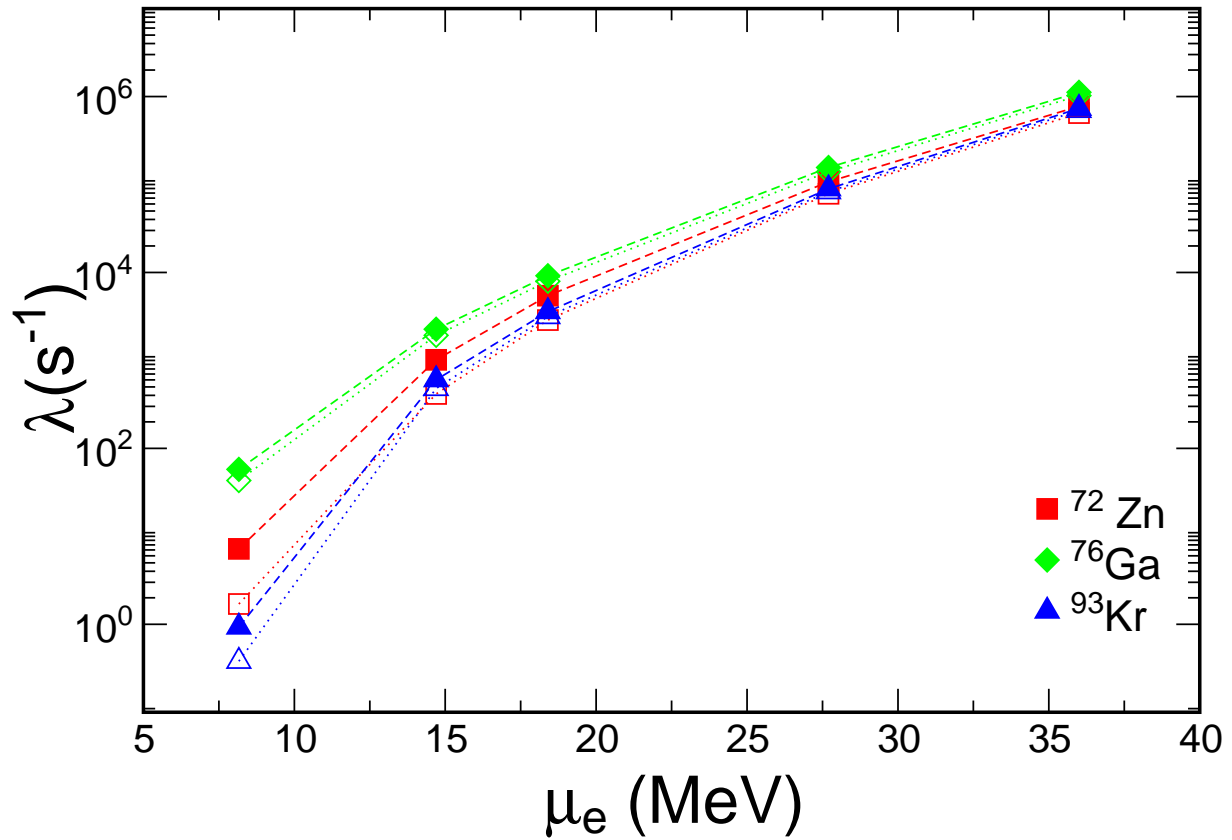}
\end{center}
\caption{
Dependence of the electron capture rates 
($\lambda$) on the $\mu_n-\mu_p$ difference 
for five relevant chemical potentials 
($\mu_e$) along the collapse trajectory. Filled points represent the 
rates for temperature-independent $\mu_n-\mu_p$ differences, and non-filled 
points represent the rates for the temperature-dependent differences.}
\label{traj}
\end{figure}

We show the results in Fig.~\ref{traj} as a function of the electron
chemical potential $\mu_e$ in the environment, calculated at five points along
a typical collapse
trajectory made available to us by \cite{Liebendoerfer}. 
Note that $\mu_e$ scales with the density as $\sim \rho^{1/3}$.
At the lowest density point ($\rho \approx 10^{10}$ g/cm$^3$), the
electron chemical potential $\mu_e$ is of the same order as the Q-value
for $^{72}$Zn (a representative nucleus at these conditions), making
the capture rate sensitive to a change in $\mu_n-\mu_p$ difference. We
find that the capture rate is decreased by a factor of order 3, if the
difference increases as suggested by \cite{Donati}. $^{76}$Ga, 
another nucleus present in the matter composition at these conditions,
has a smaller Q-value. Hence, the effect of an increase in $\mu_n-\mu_p$
does affect the rate significantly less. This shows the
importance which an increase of the symmetry energy with temperature 
in the capture rates rests on
the competition between the growth of the electron chemical potential
and the typical Q-values of the nuclei in the matter composition
along the collapse trajectory. At about $10^{11}$ g/cm$^3$, $\mu_e$ is
of order 20 MeV, while typical Q-values are of order 10 MeV. Thus, an
effective change in the $Q$-value by about $10\%$ has only little effect
on the capture rate. This can be seen in Fig. \ref{traj} 
where the effect of a temperature-dependent increase in the
$\mu_n-\mu_p$ difference becomes less relevant with increasing density.
This suggests that the proposed temperature dependence of
the symmetry energy decreases the capture rate on nuclei during a small
period of the collapse. The absolute changes, even during this period,
appear to be rather mild so that one does not expect significant changes
for the collapse trajectory.

In conclusion, we performed SMMC calculations in an 
$0f$-$1p$-$0g_{9/2}$-$1d_{5/2}$ model space using an 
effective pairing+quandrupole Hamiltonian. 
We found that the symmetry energy increases as a function of 
temperature by about 6\% on average. When applying this symmetry 
change to electron captures occurring along a typical core-collapse mass
trajectory, we found marginal differences in the calculated rates.

\acknowledgements
Our work was supported by the Danish Research Council. 
J.M.S. acknowledges the financial support of the 
Portuguese Foundation for Science and Technology. Oak Ridge National 
Laboratory is operated by UT-Battelle, LLC
for the U.S. Department of Energy under contract DE-AC05-00OR22725. 

%\pacs{PACS numbers: 26.50.+x, 23.40.-s, 21.60.Cs}

\end{document}